\documentclass{aa}
\usepackage{graphicx}
\usepackage{txfonts}
\begin{document}

\title{Oxygen, neon, and iron X-ray absorption in the local interstellar medium}

\author{Efra\'in~Gatuzz\inst{1,2,3},
        Javier~A.~Garc\'ia\inst{4},
        Timothy~R.~Kallman\inst{5},
        \and
        Claudio~Mendoza\inst{3}
        }

   \institute{Max-Planck-Institut f\"ur Astrophysik,
85741 Garching bei M\"unchen, Germany
\email{egatuzz@mpa-garching.mpg.de}
         \and
             Escuela de F\'isica, Facultad de Ciencias, Universidad
Central de Venezuela, PO Box 20632, Caracas 1020A, Venezuela
         \and
             Centro de F\'isica, Instituto Venezolano de Investigaciones
Cient\'ificas (IVIC), PO Box 20632, Caracas 1020A,  Venezuela
\email{claudio@ivic.gob.ve}
              \and
              Harvard-Smithsonian Center for Astrophysics, Cambridge, MA, 02138,  USA
\email{javier@head.cfa.harvard.edu}
\and
NASA Goddard Space Flight Center, Greenbelt, MD 20771, USA
\email{timothy.r.kallman@nasa.gov}
             }

     \abstract
  % context heading (optional)
  % {} leave it empty if necessary
   {}
  % aims heading (mandatory)
   {We present a detailed study of X-ray absorption in the local interstellar medium by analyzing the X-ray spectra of 24 Galactic sources  obtained with the {\it Chandra} High Energy Transmission Grating Spectrometer and the {\it XMM-Newton} Reflection Grating Spectrometer.}
  % methods heading (mandatory)
   {By modeling the continuum with a simple broken power law and by implementing the new {\tt ISMabs} X-ray absorption model, we estimated the total H, O, Ne, and Fe column densities toward the observed sources.}
  % results heading (mandatory)
   {We have determined the absorbing material distribution as a function of source distance and galactic latitude--longitude.}
  % conclusions heading (optional), leave it empty if necessary
   {Direct estimates of the fractions of neutrally, singly, and doubly ionized species of O, Ne, and Fe reveal the dominance of the cold component, thus indicating an overall low degree of ionization.  Our results are expected to be sensitive to the model used to describe the continuum in all sources.}


   \maketitle
%
%________________________________________________________________

%%%%%%%%%%%%%%%%%%%%%%%%%%%%%%%%%%%%%%%%%%%%%%%%%%%%%%%%%%%%%%%%%%%%%%%%%%%%%%%
\section{Introduction}\label{sec_intro}

The interstellar medium (ISM) is one of the most important galactic components. Regarding its chemical composition, it can be enriched with heavy elements by gas accretion from other galaxies, supernova explosions, and stellar winds \citep{pin13}. The analysis of such a dynamic environment is crucial for understanding stellar formation and evolutionary processes, which may be performed by means of high-resolution X-ray spectroscopy. Owing to their high penetrating power, X-ray photons interact not only with the ISM atomic ions exciting inner-shell levels but also with molecules and solid compounds, thus providing the opportunity to study physical properties, such as column densities, ionization fractions, and abundances of astrophysically relevant elements in both the gas and grains \citep{gat15}.
%-+

Multiple ISM analyses using low-mass X-ray binaries (LMXB) have been carried out in the past decade that report the presence of a multiphase ISM structure, which includes a cold gas with low-ionization degree and a hot ionized gas \citep[e.g.,][]{sch02, tak02, jue04, yao09, lia13, pin13}; however, the identification of the highly ionized species is less straightforward than the lowly ionized ones. For example, \citet{gat14} performed a study of eight LMXB using high-resolution {\it Chandra} spectra, detecting only a K$\alpha$ absorption feature ($22.024\pm 0.003$~\AA) from highly ionized \ion{O}{vi} toward XTE~J1817-330. This line was previously identified by \citet{gat13b, gat13a} as being intrinsic to the source rather than originating in the local ISM. They concluded that the cold phase is dominant without any further identification of an associated hot gas component. In support of this view, \citet{luo14} have found that most of the absorption lines from highly ionized metals detected in the spectra toward 12 LMXB arise in the hot gas intrinsic to the sources, the ISM only making a small contribution. In a recent study, \citet{nic14} has concluded that the absorption line at $\sim 22.275$~\AA\ is ubiquitously detected in the spectra of both Galactic and extragalactic sources. It leads to identifying \ion{O}{ii} K$\beta$ from the local ISM, thus contradicting its previous identification as an \ion{O}{vi} K$\alpha$ absorption feature of the warm-hot intergalactic medium (WHIM) along the lines of sight toward H~2356--309 \citep{buo09,fan10} and Mkn~501 \citep{ren14}. %-+

%%%%%%%%%%%%%%%%% Aitoff proyection figure  %%%%%%%%%%%%%%%%%%
   \begin{figure*}
   \resizebox{\hsize}{!}{\includegraphics{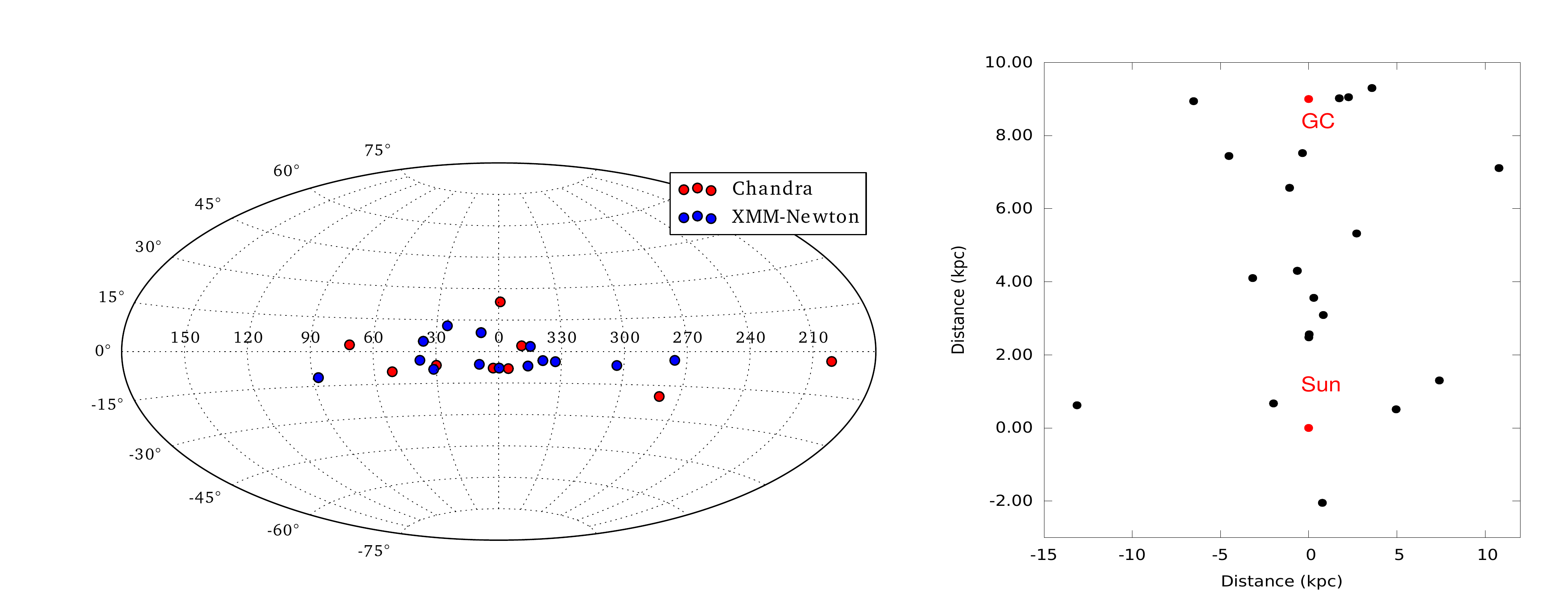}}
      \caption{Location map of the X-ray binaries of the present sample. The left panel shows the sources plotted on a full-sky map in galactic coordinates relative to the Galactic center. Red points: {\it Chandra} sources. Blue points: {\it XMM-Newton} sources. The right panel shows a location map of the X-ray binaries in our sample projected in the Galactic plane. }\label{fig1}
   \end{figure*}

%%%%%%%%%%%%%%%%%%%%%%%%%%%%%%%%%%%%%%%%%%%%%%%%%%%%%%%%%%%%%%%%%%%%%%%%%%%%%%%

\citet{gat13b,gat13a} made a detailed evaluation of the oxygen absorption region using {\it Chandra} high-resolution spectra from the bright binary XTE~J1817--330, observing a complex structure around the K edge that included the presence of K$\alpha$, K$\beta$, and K$\gamma$ absorption lines from \ion{O}{i} and \ion{O}{ii} and K$\alpha$ from \ion{O}{iii}, \ion{O}{vi}, and \ion{O}{vii}. \citet{gat14} fit the O K edge in the {\it Chandra} spectra of eight LMXB with the {\tt warmabs} photoionization model to determine the ISM ionization degree. In the same work, the O-photoabsorption cross-section computed by \citet{gar05} was compared with the most recent calculation by \citet{gor13}, finding no substantial differences in the modeling of the K-shell region. It is worth emphasizing the remarkable accuracy of the oxygen atomic data that was thereby established. The dominance of a neutral gas in all the observed lines of sight  was therefore concluded in this work with ionic fractions \ion{O}{ii}/\ion{O}{i} and \ion{O}{iii}/\ion{O}{i} lower than 0.1.
%-+

In addition to oxygen, a benchmark of the neon atomic data was also pursued by \citet{gat15}. Using the high signal-to-noise spectra of Cygnus~X--2 and XTE~J1817--330, the positions of the K$\beta$ absorption line in \ion{Ne}{i} and K$\alpha$, K$\beta$, and K$\gamma$ in \ion{Ne}{ii} and \ion{Ne}{iii} were determined. The \ion{Ne}{i} cross-section computed by \citet{gor00} and those for \ion{Ne}{ii} and \ion{Ne}{iii} by \citet{jue06} were adjusted by shifting the wavelength scale in order to fit the astronomical spectra. With the improved cross sections, a new X-ray absorption model, referred to as {\tt ISMabs}, was developed and made publicly available \citep{gat15}. Its advantage lies in the compilation of a database of photoabsorption cross sections for neutrally, singly, and doubly ionized species for all the cosmic abundant elements, namely H, He, C, N, O, Ne, Mg, Si, S, Ar, Ca, and Fe. Although the neutral component is dominant in the ISM, it is found that including singly and doubly ionized species leads to more realistic models of the O and Ne K edges; furthermore, the use of a physical model prevents misidentification of absorption features that often occurs in the traditional fitting methods with Gaussian profiles.
%-+

ISM X-ray absorption affects all the X-ray observations and is particularly evident in high-resolution grating spectra. Even though a careful modeling of the ISM was previously attempted, its chemical composition and spatial distribution are still being debated (e.g., identification of a Galactic highly ionized gas and the existence of a homogeneous distribution on large scales). Moreover, an imprecise ISM modeling can lead to erroneous conclusions when analyzing other environments, such as the Galactic halo or the WHIM. In this respect, a comprehensive analysis of the ISM through multiple lines of sight and use of a realistic absorption model like {\tt ISMabs} would be invaluable for addressing these concerns; consequently, we present an extensive analysis of the ISM using {\it XMM-Newton} and {\it Chandra} high-resolution X-ray spectra. 

The outline of this report is as follows. In Section~\ref{sec_dat} the data reduction process is summarized, and in Section~\ref{sec_broad} the data-fitting procedure is described. A thorough discussion of the results is given in Section~\ref{sec_dis}, and finally in Section~\ref{sec_con}, we draw the conclusions of this work.
%-+

%%%%%%%%%%%%%%%%%%%%%%%%%%%%%%%%%%%%%%%%%%%%%%%%%%%%%%%%%%%%%%%%%%%%%%%%%%%%%%%
\section{Observations and data reduction}\label{sec_dat}

To carry out the present ISM study we have gathered a sample of X-ray binary spectra from the {\it XMM-Newton} Science Archive (XSA\footnote{ http://xmm.esac.esa.int/xsa/}) and the {\it Chandra} Source Catalog (CSC\footnote{ http://cxc.harvard.edu/csc/}). A total of 24 bright sources were analyzed, 15 from the XSA and 17 from the CSC. Some of them have been observed with both telescopes; details of the observations are reviewed in the Appendix~\ref{sec_apx_fit}. As shown in Figure~\ref{fig1}, the source locations allow an analysis of the ISM along different lines of sight, and it also depicts the source position projected in the Galactic plane.  The {\it XMM-Newton} spectra were reduced with the
standard Scientific Analysis System (SAS) threads\footnote{
http://www.cosmos.esa.int/web/xmm-newton/sas-threads/} and the {\it
Chandra} spectra with the standard CIAO threads\footnote{http://cxc.harvard.edu/ciao/threads/gspec.html}. We estimated the zero-order position for {\it Chandra} spectra with the {\tt findzo} algorithm\footnote{http://space.mit.edu/cxc/analysis/findzo/}. All spectra were rebinned to 25 counts per channel in order to use $\chi^2$ statistics. The spectral fitting was performed with the {\sc isis} data analysis package \citep[version 1.6.2-30\footnote{http://space.mit.edu/asc/isis/}]{hou00}.
%-+

%%%%%%%%%%%%%%%%%%%%%%%%%%%%

\begin{table*}
\tiny
\caption{\label{tab3}{\tt ISMabs} column density values.}
\centering
\begin{tabular}{lccccccccc}
\hline\hline
Source&\multicolumn{8}{c}{{\tt ISMabs}}\\
\cline{2-9}\\
 & $N({\rm H}) $ & $N({\rm O~{\sc I}})$& $N({\rm O~{\sc II}})$& $N({\rm O~{\sc III}})$& $N({\rm Ne~{\sc I}})$& $N({\rm Ne~{\sc II}})$& $N({\rm Ne~{\sc III}})$& $N({\rm Fe})$ \\
\hline
4U~0614+091& $9.3 \pm 0.7$   &  $17.9  \pm 2.3$   & $<0.57 $ & -- &  $4.6  \pm    0.7 $  & $ 0.25  \pm     0.17$   & $<0.02 $ &  $0.49  \pm     0.11$  \\
4U~0918--54& $<1.5 $ & $17.7\pm 1.7$ & $<0.01 $ &  $<0.2$ &  $2.3  \pm 1.3$   &  $1.09  \pm     1.04$   &$ <0.01 $ &  $0.72  \pm     0.19$  \\
4U~1254--69& $2.2 \pm 0.2$ & $14.6 \pm 0.8$ & $0.28 \pm 0.24$ & $0.34 \pm 0.16$ & $2.0 \pm 0.3$ & $0.79 \pm 0.25$ & $0.05 \pm 0.04$ &$0.59\pm  0.06$  \\
4U~1636--53 & $5.6 \pm 1.7$ & $22.9 \pm 4.3$ &  $0.72  \pm 0.32$ & $0.35  \pm  0.14$ & $3.2 \pm  0.9$ & $0.78 \pm 0.40 $& $<0.06 $ & $1.05 \pm  0.22 $\\
4U~1728--16 & $6.6 \pm 0.4$ & $14.2 \pm 1.2$ & $0.30 \pm 0.18$ & $<0.16$ & $3.5 \pm 0.4$ & $0.54 \pm 0.23 $ & $<0.06$ & $0.48 \pm 0.06$  \\
4U~1735--44 & $ 3.2 \pm 1.3$ & $12.1 \pm 1.7$ &$0.53 \pm 0.23$& $0.23 \pm 0.10$ & $2.9 \pm 1.3$ &$0.61 \pm 0.42$ & $<0.16$ & $0.76 \pm  0.21$ \\
4U~1820--30 &$ 2.4 \pm 0.8$ & $8.9 \pm  2.2$ & $<0.70 $ & $<0.48 $ & $1.0 \pm  0.7 $& $0.16 \pm  0.15$ & $<0.02 $ & $0.50 \pm  0.15$ \\
4U~1915--05 &$4.7 \pm 2.1$& $23.3 \pm 4.3 $ & $<0.02$& -- & $4.4 \pm 1.9$ & $<0.01 $ & -- & $1.15 \pm 0.48$  \\
4U~1957+11 &$  3.4  \pm     1.7$   & -- & -- & -- & $ 1.4  \pm     1.6$   & $<0.16 $ &$ <0.01 $ &$  0.67  \pm     0.28$  \\
Aql~X--1 &$ 3.6 \pm 0.3$ &$23.6 \pm 1.2$ & $ < 0.06$ & $0.21 \pm 0.13$ & $5.3 \pm 0.3$ &$0.58 \pm 0.20$ &$<0.04$ & $ 1.13 \pm  0.07$  \\
Cygnus~X--1& $9.7 \pm 0.4$ & $ 21.4 \pm 1.4$ & $0.28 \pm 0.18$ &$0.47 \pm 0.16$ & $6.6 \pm 0.3$ & $0.39 \pm 0.09$ & $0.05 \pm 0.02$ & $0.98 \pm 0.04$  \\
Cygnus~X--2 &$4.3 \pm 0.4$ &$11.4 \pm 0.8$ & $0.45 \pm 0.19 $& $0.15 \pm  0.08$ & $2.6 \pm  0.2$ & $0.50 \pm 0.10$ &$0.06 \pm  0.03$ & $0.44 \pm 0.05$ \\
EXO~0748--676 & $<2.5 $& -- & -- & -- & $ 6.1  \pm     2.8 $  & -- & -- & -- \\
GRO~J1655--40 & $7.8 \pm 0.3 $&$26.2 \pm 1.4 $&$2.55 \pm 1.08$ & $3.37 \pm 1.10$ & $7.7 \pm 0.3 $& $1.11 \pm 0.19 $& $<0.06$ & $1.61 \pm 0.13$ \\
GS~1826--238 & $3.1 \pm 0.3$ & $ 23.3 \pm 1.5 $  &$0.28 \pm  0.22$ & $0.74 \pm 0.25$ & $5.0 \pm 0.4$ &$0.73 \pm 0.28$ &$ <0.01$ & $ 1.07 \pm  0.07$  \\
GX~339--4 &$4.1 \pm 0.5$ & $30.2 \pm 3.1$ & $<1.79$ & $<1.19$ & $4.9 \pm 0.5$ & $1.72 \pm  0.36 $& $0.19 \pm  0.08 $& $1.57 \pm  0.17$ \\
GX~349+2 &$ 14.3 \pm  3.3$ & -- & -- & -- & $11.0 \pm  2.2$ & $2.11 \pm  1.54 $&$ <0.20 $ & $2.31 \pm  0.99 $\\
GX~9+9  &$7.4 \pm 0.7$ & $15.0 \pm 1.7 $& $0.36 \pm 0.16 $  & $ <0.04 $  & $3.2 \pm  0.4$ & $0.60 \pm  0.19$ & $<0.05 $ & $0.52 \pm  0.09 $\\
J1753.5--0127 &$ <0.4 $& $ 18.2 \pm 1.4$&$ <0.22 $ & $ 0.53 \pm 0.49 $  &$2.3 \pm 0.8$& $ <0.15$ & $<0.06$& $0.53  \pm  0.18 $ \\
Swift~J1808-3658 & $ 2.2  \pm     1.3 $  &  $7.0  \pm     3.9  $ & -- & -- & -- & -- & -- & -- \\
Swift~J1910.2-0546 & $7.4 \pm 1.2$& $ 12.9\pm 3.8$ & $<0.92$ & -- & $3.6  \pm 0.6$ &  $0.42 \pm  0.17$  &  $0.09 \pm  0.04$   &  $0.91  \pm     0.14$  \\
Sco~X--1 & $ 2.1  \pm     1.5$   & $ 9.4  \pm    1.4 $  & $<0.11 $ &$ <0.01 $ & -- & -- & -- & $ 0.42 \pm  0.12 $ \\
Ser~X--1 & $3.0 \pm  0.9$ & $32.5 \pm  3.0 $&$ <1.08 $ &$  0.52  \pm     0.29 $  &$ 6.0 \pm  0.7$ & $1.04\pm  0.44$ &$ <0.06 $ & $1.46 \pm  0.21$ \\
XTE~J1817-330 & $1.4 \pm  0.4 $&$ 11.1 \pm  0.5$ & $0.72 \pm  0.24$ & $0.34 \pm  0.12$ & $1.4 \pm  0.2$ & $0.41 \pm  0.10$ & $0.11 \pm  0.04 $& $0.46 \pm  0.03$ \\
\hline
\end{tabular}
\tablefoot{$N({\rm H})$ in  units of $10^{21}$~cm$^{-2}$ and metal column densities in units of $10^{17}$~cm$^{-2}$.}
\end{table*}

%%%%%%%%%% O Chandra %%%%%%%
   \begin{figure*}
     \resizebox{\hsize}{!}{\includegraphics{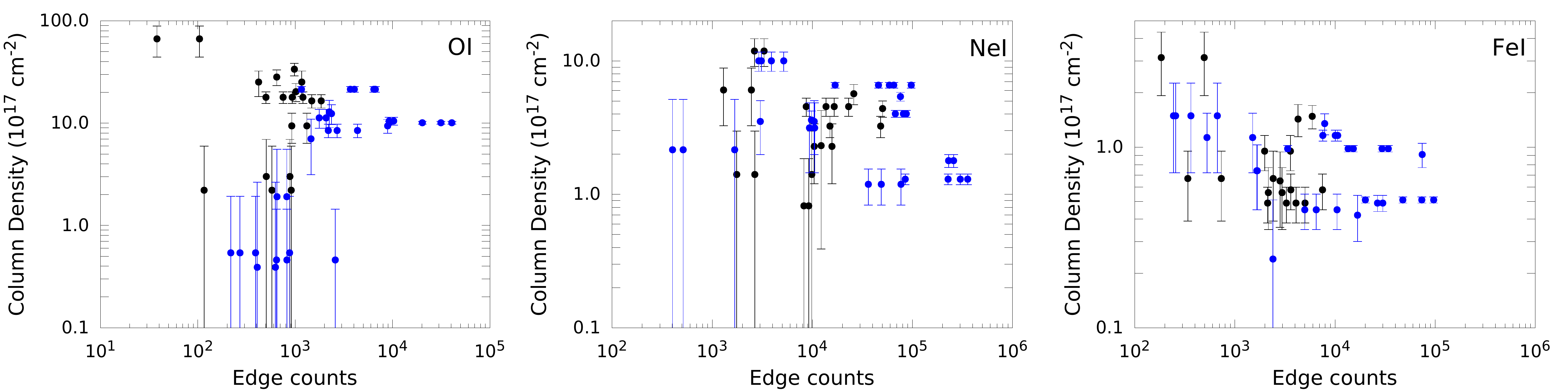}}
     \caption{{\tt ISMabs} column densities for {\it Chandra} observations as a function of the number of counts. Black points correspond to {\it Chandra} TE-mode spectra and blue points to {\it Chandra} CC-mode spectra.}\label{fig3}
   \end{figure*}

%%%%%%%%%%%%%%%%%%%%%%%%%%%%%%%%%%%%%%%%%%%%%%%%%%%%%%%%%%%%%%%%%%%%%%%%%%%%%%%
\section{Spectral fitting procedure}\label{sec_broad}

To estimate the O, Ne, and Fe column densities we carried out a broadband fit (11--24~\AA) for each source listed in the Appendix~\ref{sec_apx_fit} using a simple {\tt ISMabs*Bknpower} model. The {\tt Bknpower} component corresponds to a broken power law that in general provides a better fit to the continuum than a single unbroken power law. In the case of Sco~X--1, we analyzed the spectra in the 15--24~\AA\ region because of the absence of data below this wavelength range. The {\tt ISMabs} model \citep{gat15} includes photoabsorption cross sections for neutrally, singly, and doubly ionized species, and the Fe photoabsorption cross-section is taken from  metallic iron laboratory measurements \citep{kor00}. For each source all observations were fit simultaneously (i.e., using the same absorption parameters and varying the normalization for each
observation); in the case of {\it Chandra}, we considered observations with timed-exposure (TE) readout mode and continuous clocking (CC) readout mode separately. The free parameters for the {\tt ISMabs} fits were the H, O, Ne, and Fe column
densities, including the \ion{O}{ii}, \ion{O}{iii}, \ion{Ne}{ii}, and \ion{Ne}{iii} ions.
%-+

Data statistics (e.g., number of counts in the wavelength region of interest) clearly have a significant impact on the absorption modeling. Figure~\ref{fig3} shows the {\tt ISMabs} column densities obtained from the {\it Chandra} observations as a function of the number of counts in the O  (21--24~\AA), Ne  (13--15~\AA), and Fe  (16--18~\AA) absorption regions, which brings out a disparity in the O columns when the number of counts is low. This is probably because the details of the spatial distribution are lost
in the CC read mode, and there is therefore no way to spatially separate the background from the source. This affects the spectra at long wavelengths where the effective area is smaller and where the background may become comparable to the source signal.
%-+

Specifically, CC-mode observations with less than approximately 2000 counts in the O K-edge region give rise to the outliers in the lefthand panel of Figure~\ref{fig3}. Therefore, in the following procedures, we excluded observations with a limited number of counts in the O K-edge region (21--24~\AA) and Ne K-edge region (13--15~\AA). Additional information on the influence of {\it Chandra} readout modes in high-resolution spectroscopy are given in Appendix~\ref{sec_apx_cc}. Finally, for those sources with observations from both {\it Chandra} and {\it XMM-Newton}, we have found that the derived columns are in good agreement, so we take the average of the fit values.
%-+

%%%%%%%%%%%%%%%%%%%%%%%%%%%%%%%%%%%%%%%%%%%%%%%%%%%%%%%%%%%%%%%%%%%%%%%%%%%%%%%
\section{Results and discussions}\label{sec_dis}

\begin{figure*}
   \resizebox{\hsize}{!}{\includegraphics{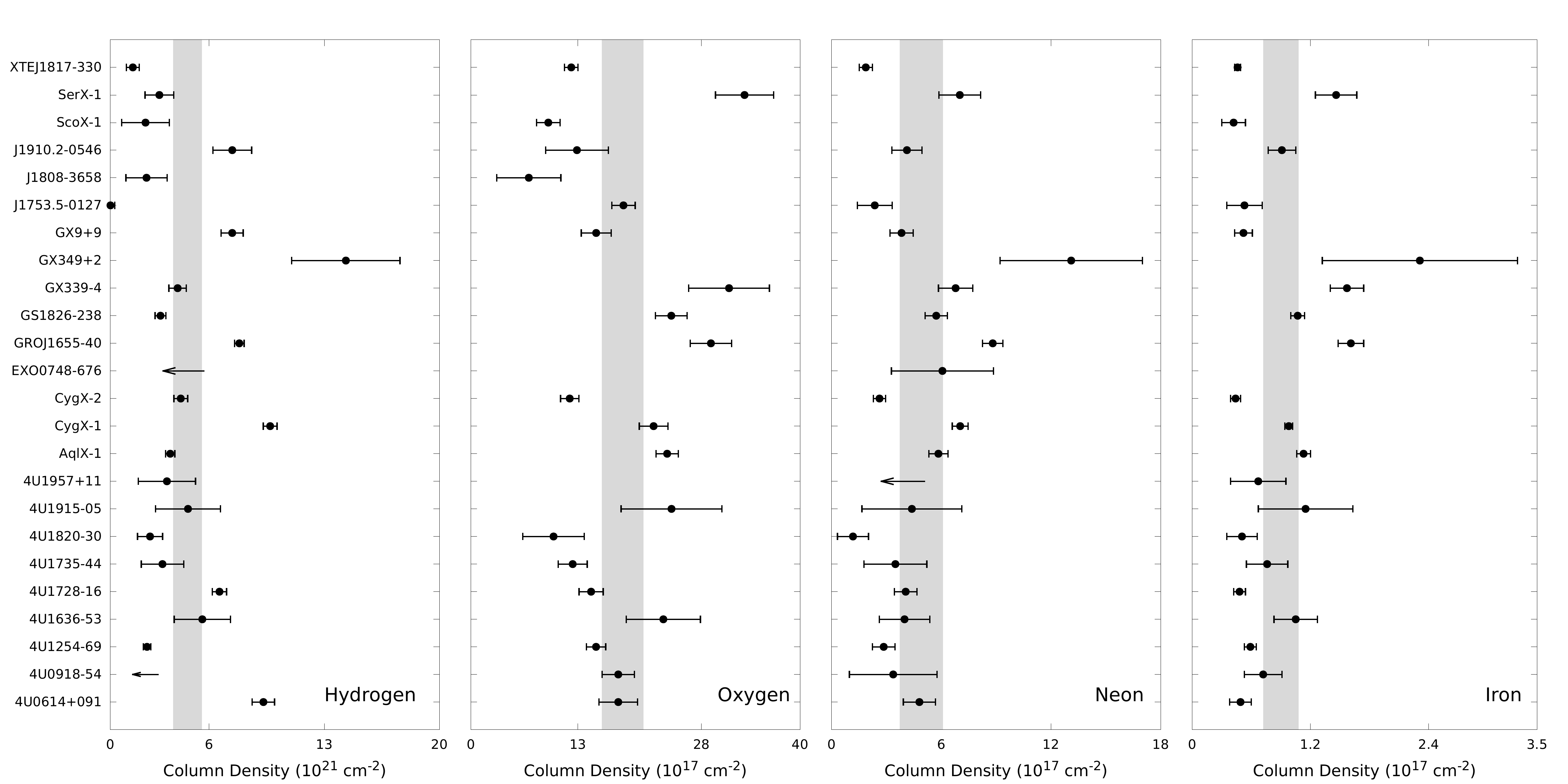}}
      \caption{{\tt ISMabs} column densities values for each source.  Gray boxes correspond to a 2$\sigma$ region around the average value.}\label{fig4}
   \end{figure*}

The fit results to all the observations in our sample are listed in Table~\ref{tab3}.  Upper limits for the column densities (i.e., the highest values obtained for each ion considering their upper errors) in units of $10^{17}$~cm$^{-2}$ are $N({\rm O~{\sc I}})< 32.46$; $N({\rm O~{\sc II}})< 2.55$; $N({\rm O~{\sc III}})< 3.37$; $N({\rm Ne~{\sc I}})< 10.95$; $N({\rm Ne~{\sc II}})< 2.11$; $N({\rm Ne~{\sc III}})< 0.20$; and $N({\rm Fe})< 2.31$. Figure~\ref{fig4} shows the column density distribution for each source. The hydrogen and neon column densities are mostly consistent and are homogeneously distributed along their average value; this is expected in neon since it does not form molecules.  Oxygen and iron column densities, on the other hand, tend to be more dispersed along the different lines of sight. This is discussed further in Section~\ref{sec_apx_mol}.
%-+

%%%%%%%%%% Column densities for each source %%%%%%%

\begin{table}
\caption{\label{tab4}Hydrogen column density comparison.}
\small
\centering
\begin{tabular}{lccccccccc}
\hline
Source&N({\rm H})&$21$~cm&$21$~cm&$21$~cm\\
&&survey$^a$&survey$^b$&survey$^c$\\
\hline
4U~0614+091 &  $9.30 \pm 0.68$   &5.42 & 4.42 & 5.86 \\
4U~0918--54 & $<1.53 $ & 7.34 & 6.19 & 7.63    \\
4U~1254--69 & $2.23  \pm     0.23$ &  2.85 & 2.15 & 3.46    \\
4U~1636--53 & $5.59 \pm  1.71$ &3.30 & 2.64 &4.04 \\
4U~1728--16 & $6.63  \pm     0.43$ & 2.10 & 1.98 & 3.31    \\
4U~1735--44 & $ 3.17 \pm  1.29$ &3.03 & 2.56 &3.96 \\
4U~1820--30 &$ 2.41 \pm  0.76$ &1.52 & 1.32 &2.33 \\
4U~1915--05 &$ 4.72  \pm     1.97 $&2.55 & 2.31& 3.72    \\
4U~1957+11 &$  3.44  \pm     1.74$  &1.31 & 1.23 & 2.01 \\
Aql~X--1 &$ 3.64  \pm     0.28$ &3.28 & 2.86 & 4.30     \\
Cygnus~X--1 & $ 9.71  \pm     0.42 $ &8.10 & 7.81 & 9.25 \\
Cygnus~X--2 &$ 4.28 \pm  0.42$ &2.20 & 1.88 & 3.09  \\
EXO~0748--676 & $<2.47 $&2.03 & 1.87 & 3.21 \\
GRO~J1655--40 & $7.84 \pm  0.29 $&6.84 & 5.78 & 7.22 \\
GS~1826--238 & $3.05  \pm     0.33$ & 1.83 & 1.68 & 3.00    \\
GX~339--4 &$ 4.09 \pm  0.53$ &5.00 & 3.74 &5.18   \\
GX~349+2 &$ 14.31 \pm  3.29$ &6.12 & 4.69 & 6.13 \\
GX~9+9  &$ 7.40 \pm  0.67$ &2.10 & 1.98 & 3.31 \\
J1753.5--0127 &$ 0.01  \pm     0.26 $&1.64 & 1.66 & 2.98     \\
Swift~J1808-3658 & $ 2.20  \pm     1.25 $  &0.27&0.26&0.29 \\
Swift~J1910.2-0546 & $ 7.41  \pm     1.18$  &2.24       &2.44    &2.24   \\
Sco~X--1 & $ 2.14  \pm     1.45$   &1.47 & 1.40 & 2.54 \\
Ser~X--1 & $2.98 \pm  0.87$ &4.43 & 3.98 & 5.42 \\
XTE~J1817-330 & $1.37 \pm  0.39 $&1.58 & 1.39 & 2.29 \\
\hline
\end{tabular}
\tablefoot{$N({\rm H})$ in units of $10^{21}$~cm$^{-2}$. $^a$\citet{dic90}; $^b$\citet{kal05} and $^c$\citet{wil13}.}
\end{table}

%%%%%%%%%% N vs 21 cm %%%%%%%%%

\begin{figure}
  \centering
  \includegraphics[width=8.5cm,height=6cm]{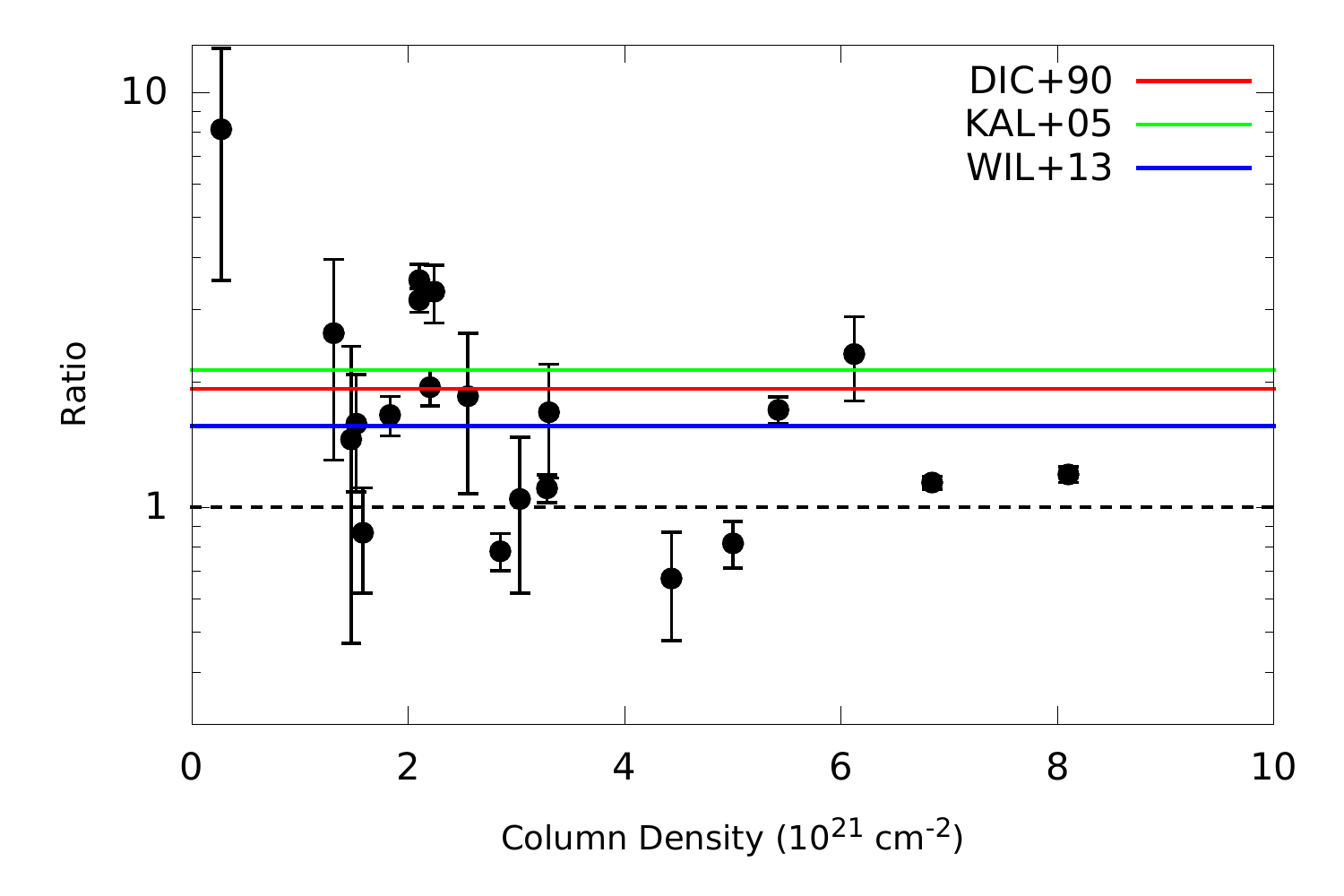}
  \caption{ Comparison of hydrogen column densities from different 21~cm surveys with the present {\tt ISMabs} results. Black data points corresponds to the {\tt ISMabs}/21~cm ratio using the \citet{dic90} measurements. The solid color lines are linear fits to the {\tt ISMabs}/21~cm values relative to the various determinations from the 21~cm surveys: \citet[red line]{dic90}, \citet[green line]{kal05}, and \citet[blue line]{wil13}.}\label{fig5}
\end{figure}

\subsection{Hydrogen column densities}\label{sec_h}

A comparison of the {\tt ISMabs} hydrogen column densities with the data sets from the 21~cm surveys \citep{dic90, kal05, wil13} is provided in Table~\ref{tab4}. The 21~cm hydrogen column densities correspond to the average value of all absorbers within $1^{\circ}$ of the source position. It is important to note that \citet{dic90} and \citet{kal05} only consider the atomic \ion{H}{i} column densit,y while \citet{wil13} include both atomic and molecular hydrogen. Figure~\ref{fig5} shows a comparison of the 21~cm measurements with those derived from our {\tt ISMabs} fits.  Black data points correspond to the {\tt ISMabs}/21~cm ratio using the \citet{dic90} measurements. The solid color lines are linear fits to the {\tt ISMabs}/21~cm values relative to the various determinations from the 21~cm surveys \citep{dic90, kal05, wil13}. For low column densities, the {\tt ISMabs} values tend to be systematically higher than the 21~cm values, while for high column densities, the {\tt ISMabs} values tend to be lower than the 21~cm values. We found overall that the best agreement of our results is with those from \citet{wil13}.

%%%%%%%%%% N vs Distances/Latitude/Longitude %%%%%%%
\begin{figure*}
  \resizebox{\hsize}{!}{\includegraphics{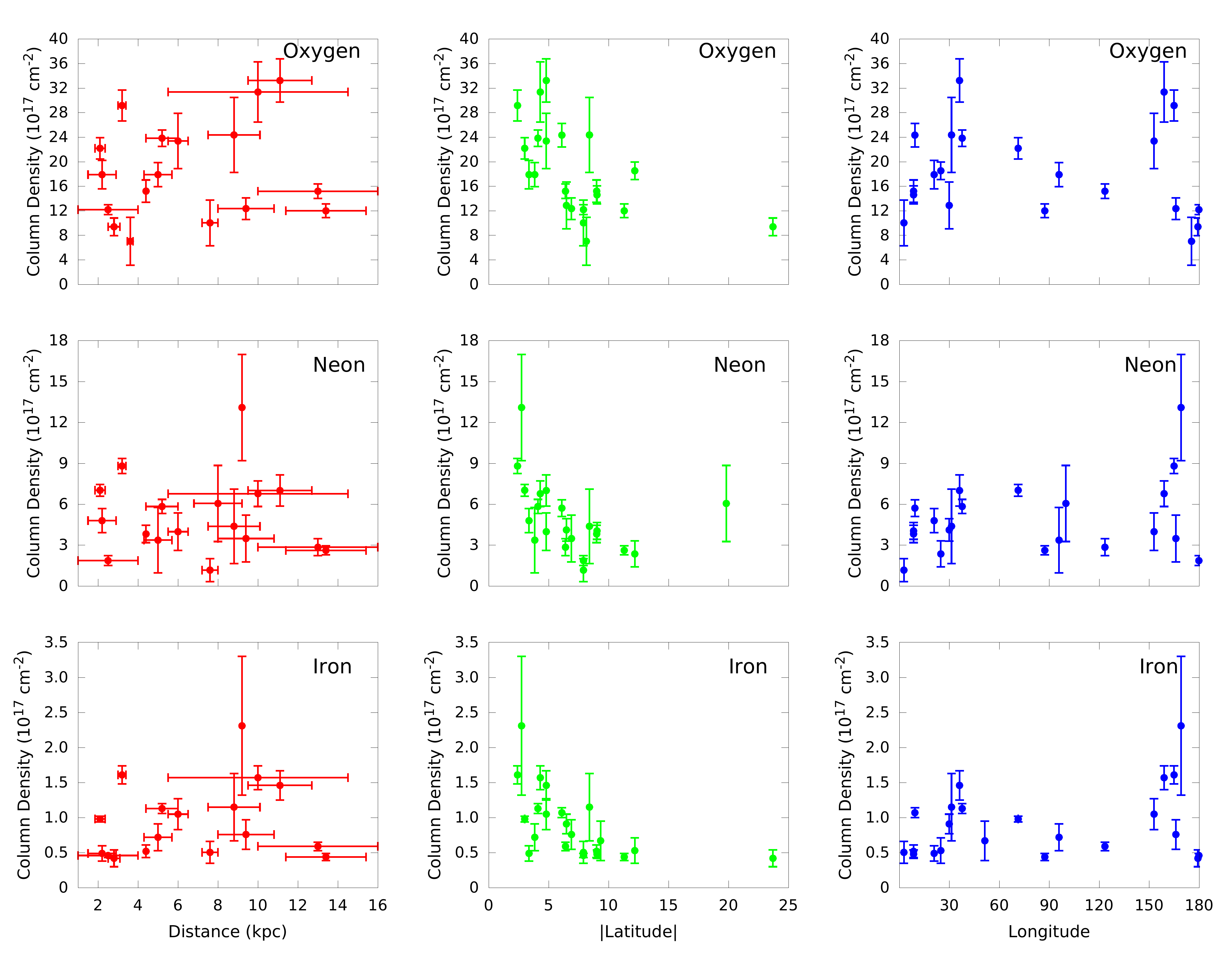}}
  \caption{Comparison of the oxygen, neon, and iron total column densities obtained from the {\tt ISMabs} model as function of distance (left panels), latitude absolute value (middle panels), and longitude (right panels) for all the analyzed sources. Longitude has been rescaled to $0^{\circ}{-}180^{\circ}$.}\label{fig6}
\end{figure*}

These discrepancies may arise from either a real difference in the intrinsic absorption or the continuum model; also, changes in the region around the X-ray source due to changes in the intrinsic column (e.g., the presence of winds) can increase absorption, and they are thus not resolved by the 21~cm surveys. These effects are more likely to be important in sources with strong stellar winds, such as Cyg~X--1 and Sco~X--1 \citep{gat14}. On the other hand, the hydrogen column densities obtained from broadband fits are subject to the uncertainty of the assumed underlying continuum. In some cases, the characterization of the spectra with different models, such as {\tt Powerlaw}, {\tt Bknpower}, and {\tt Blackbody,} leads to differences in the column densities despite similar statistics.  Therefore, since we are not able to favor a continuum model based on fit statistics, we have adopted the {\tt Bknpower} model for all observations for consistency and simplicity. We emphasize that the continuum model must be treated carefully in any attempt to perform an ISM analysis with X-ray high-resolution spectroscopy. In this respect our results are expected to be sensitive to the particular model used to describe the continuum in all sources.
%-+

\begin{table}
\caption{\label{tab5}Abundance values.}
\small
\centering
\begin{tabular}{lccccccccc}
\hline
Source&$A_{\rm O}$&$A_{\rm Ne}$&$A_{\rm Fe}$\\
\hline
4U~0614+091 &$ 0.28 \pm 0.06$ & $0.43 \pm 0.11$ & $0.17 \pm 0.05$ \\
4U~0918--54 & $<1.73 $ & $<1.84$ & $<1.49$ \\
4U~1254--69 &$ 1.01 \pm 0.18$& $1.07 \pm 0.34$ & $0.84 \pm 0.17 $\\
4U~1636--53 & $0.62 \pm 0.31$ & $0.60 \pm 0.39$ & $0.59 \pm 0.31$ \\
4U~1728--16 & $0.33 \pm 0.05$ & $0.51 \pm 0.11$ & $0.23 \pm 0.04$ \\
4U~1735--44 & $0.58 \pm 0.32$ & $0.92 \pm 0.83$ & $0.76 \pm 0.52$ \\
4U~1820--30 & $0.61 \pm 0.42$ & $0.40 \pm 0.42$ & $0.66 \pm 0.41$ \\
4U~1915--05 & $0.76 \pm 0.51$ & $0.78 \pm 0.81$ & $0.77 \pm 0.64$ \\
4U~1957+11 & -- & $<0.38 $ & $0.62 \pm 0.57$ \\
Aql~X--1 & $0.97 \pm 0.13 $& $1.34 \pm 0.22 $& $0.98\pm 0.14$ \\
Cygnus~X--1 & $0.34 \pm 0.04$ & $0.60 \pm 0.06 $& $0.32 \pm 0.03 $\\
Cygnus~X--2 & $0.41 \pm 0.08$ & $0.51 \pm 0.11 $& $0.33 \pm 0.07 $\\
EXO~0748--676 & -- & $<2.04 $ & -- \\
GRO~J1655--40 & $0.55 \pm 0.07$ & $0.94 \pm 0.09$ &$ 0.65 \pm 0.08$ \\
GS~1826--238 & $1.18 \pm 0.22 $& $1.56 \pm 0.34$ & $1.11 \pm 0.19$ \\
GX~339--4 & $1.13 \pm 0.32$ & $1.38 \pm 0.37 $& $1.21 \pm 0.28$ \\
GX~349+2 & -- & $0.76 \pm 0.40 $& $0.51 \pm 0.34$ \\
GX~9+9  & $0.30 \pm 0.06$ &$ 0.43 \pm 0.11$ & $0.22 \pm 0.06$ \\
J1753.5--0127 & $<1.50 $ & $<1.50 $ & $<1.50 $ \\
Swift~J1808-3658 & $0.47 \pm 0.53$ & -- & -- \\
Swift~J1910.2-0546 & $0.26 \pm 0.12 $&$ 0.46 \pm 0.17$ &$ 0.39 \pm 0.12$ \\
Sco~X--1 & $0.65 \pm 0.54$ & -- & $0.62 \pm 0.60$ \\
Ser~X--1 & $1.65 \pm 0.66$ & $1.96 \pm 0.89$ &$ 1.55 \pm 0.67$ \\
XTE~J1817-330 & $1.32 \pm 0.46$ &$ 1.13 \pm 0.53$ &$ 1.06 \pm 0.37$ \\
\hline
\end{tabular}
\tablefoot{Abundances are relative to the solar values of \citet{gre98}.}
\end{table}

\subsection{ISM structure}\label{sec_val}

Figure~\ref{fig6} shows a comparison of the oxygen, neon, and iron total column densities from the {\tt ISMabs} model as a function of distance (left panels), latitude absolute value (central panels), and longitude (right panels) for all the analyzed sources. We have found that the column densities tend to increase with distance, because the slope is steeper in the case of oxygen, and the  general trend is to decrease with galactic latitude. This behavior is expected because the ISM material density decreases in the vertical direction away from the Galactic plane. In the case of galactic longitude, it is difficult to establish a clear relationship with the column density but an increase at high longitude (i.e., away from the Galactic center) is hinted.
%-+

We have derived the column-density unweighted average considering all individual sources (in units of $10^{17}$~cm$^{-2}$): $N({\rm O})=18.43\pm 2.53$, $N({\rm Ne})=4.91\pm 1.19$, and $N({\rm Fe})=0.90\pm 0.18$ . Using the total column densities for each element, we estimated elemental abundances with the relation $N_{x}=A_{x}N_{h}$, where $N_{h}$ is the hydrogen column density, $A_{x}$  the abundance, and $N_{x}$  the total column density of element $x$. The resulting values are listed in Table~\ref{tab5}. The abundance behavior, on the other hand, tends to be constant with distance and latitude, and as for longitude, sources with high hydrogen column densities tend to have increased abundances at low longitude values. Using the abundance values computed for all individual sources, we have derived unweighted average values relative to solar \citep{gre98} of $A_{\rm O}=0.70\pm 0.26$, $A_{\rm Ne}=0.87\pm 0.35$, and $A_{\rm Fe}=0.67\pm 0.28$.
%-+

\begin{figure}
  \centering
  \includegraphics[width=9cm,height=11cm]{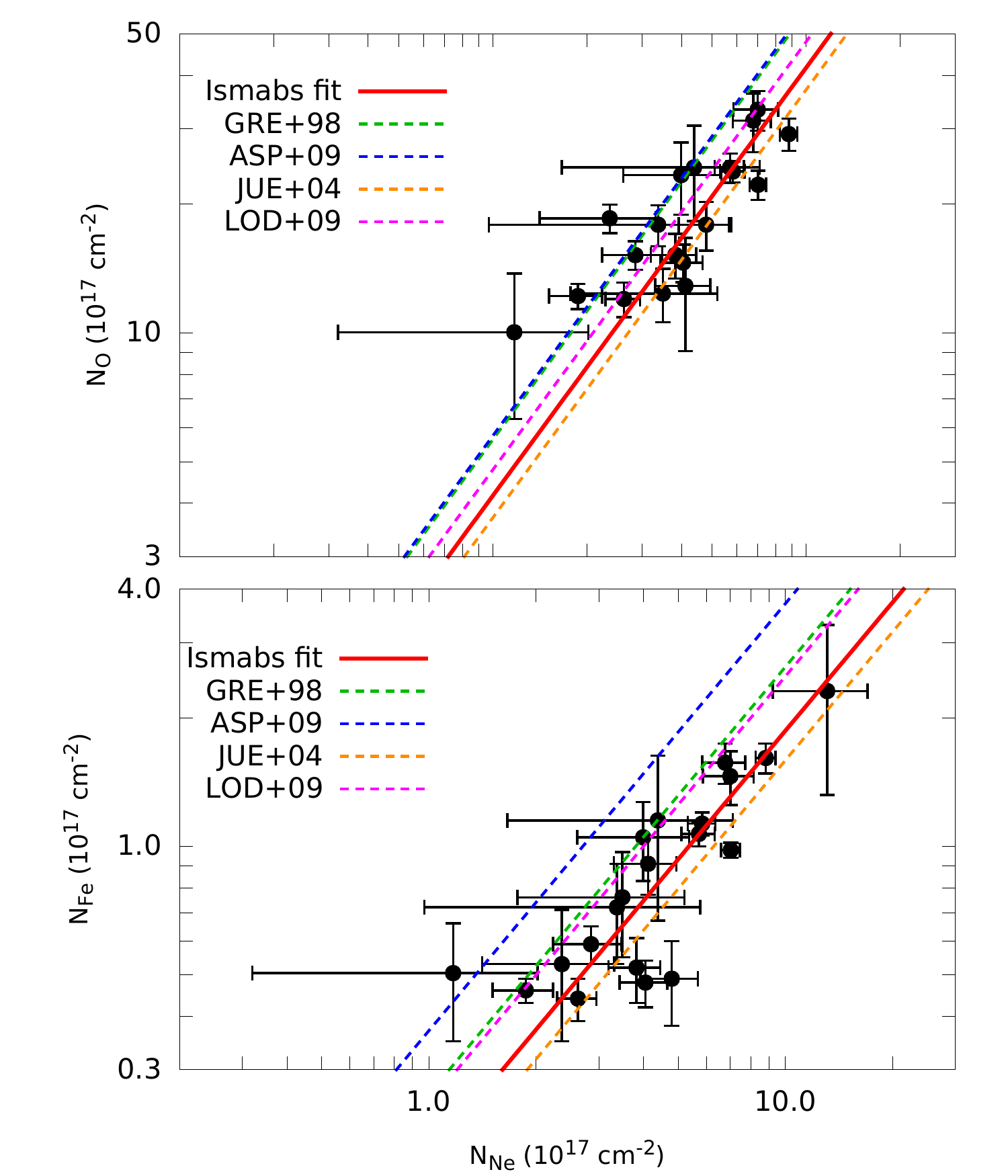}
  \caption{ Comparison of the abundance ratios $N_{\rm O}/N_{\rm Ne}$ (top panel) and $N_{\rm Fe}/N_{\rm Ne}$ (bottom panel) obtained from the broadband fit with {\tt ISMabs} for all the analyzed sources. The best-fit $N_{\rm x}/N_{\rm Ne}$ ratio for all sources is plotted with a solid red line. Ratios obtained from \citet{gre98}, \citet{asp09}, \citet{lod09}, and \citet{jue04} are respectively shown with green, blue, orange, and magenta dashed lines.}\label{fig8}
\end{figure}

The comparison of the oxygen and iron total column densities as a function of the neon column density is shown in Figure~\ref{fig8}, including contributions from the neutrally, singly, and doubly ionized species for each
element. The best-fit $N_{\rm x}/N_{\rm Ne}$ abundance ratio for all sources is plotted with the solid red line; ratios from \citet{gre98}, \citet{asp09}, \citet{lod09} and \citet{jue04} are also shown. As a noble gas, neon is a suitable baseline since it is only found in atomic form. We find an average ratio of $N({\rm O})/N({\rm Ne})=4.16\pm 0.22$, which is greater than the estimate by \citet{jue06} of $3.7\pm 0.3$ and lower than the solar values of 4.79 \citep{lod09}, $5.36$ \citep{gre98}, and $5.75$ \citep{asp09}.  We also obtain an average ratio of $N({\rm Fe})/N({\rm Ne})=0.18\pm 0.01,$ in relatively close agreement with \citet{jue06} ($0.15\pm 0.01$) and lower than the solar ratios by \citet{gre98}, \citet{lod09}, and \citet{asp09} of $0.26$, $0.25,$ and $0.37$, respectively.
%-+

\begin{figure*}
  \resizebox{\hsize}{!}{\includegraphics{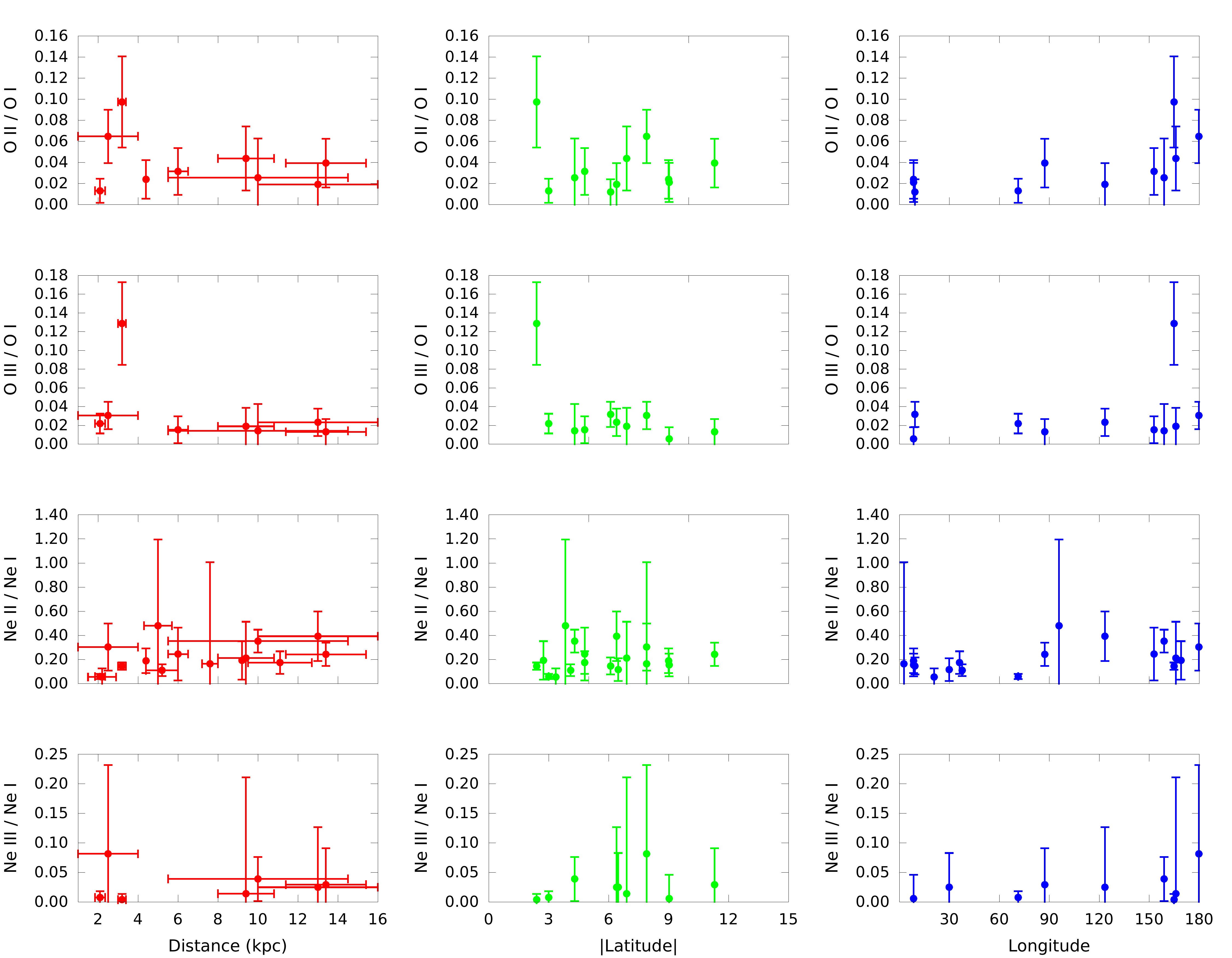}}
  \caption{Comparison of the ionization fractions \ion{O}{ii}/\ion{O}{i}, \ion{O}{iii}/\ion{O}{i}, \ion{Ne}{ii}/\ion{Ne}{i}, and \ion{Ne}{iii}/\ion{Ne}{i} obtained with the {\tt ISMabs} model as a function of distance (left panels), the latitude absolute value (middle panels), and longitude (right panels) for all these sources. Longitude has been rescaled to $0^{\circ}{-}180^{\circ}$.}\label{fig9}
\end{figure*}

The presence of ionized species in the ISM is not negligible and must be considered in order to perform a reliable chemical analysis of its environment. Figure~\ref{fig9} shows a comparison of the \ion{O}{ii}/\ion{O}{i}, \ion{O}{iii}/\ion{O}{i}, \ion{Ne}{ii}/\ion{Ne}{i}, and \ion{Ne}{iii}/\ion{Ne}{i} ionization fractions as a function of distance, latitude absolute value, and longitude.  Unlike the column densities, the ionization fractions tend to be approximately constant with the geometric parameters. This result agrees with previous findings regarding the dominance of the ISM cold phase characterized by a low ionization degree \citep{gat13b, gat13a, gat14, gat15}. We obtained the following average values:
\ion{O}{ii}/\ion{O}{i} $=0.03\pm 0.01$, \ion{O}{iii}/\ion{O}{i} $=0.03\pm 0.02$, \ion{Ne}{ii}/\ion{Ne}{i} $=0.21\pm 0.05$, and \ion{Ne}{iii}/\ion{Ne}{i} $=0.03\pm 0.02$.
%-+

%%%%%%%%%% O vs Ne column densities %%%%%%%

%%%%%%%%%%% IonizationFractions vs Distance and Ionization Fracions vs Distance/|Latitude| %%%%%%%

\subsection{Molecular absorption features}\label{sec_apx_mol}

The search for molecular and dust spectral signatures is a topic of current interest in ISM astrophysics. \citet{lee09} argue that {\it Chandra} and {\it XMM-Newton} high-resolution spectra provide a resource for constraining dust properties (e.g., distribution, composition, and abundances). Absorption features due to molecules and dust have been studied previously using LMXB \citep{dev09, kas09, pin10, cos12, pin13}; in particular, \citet{dev09} estimate an oxygen depletion rate to the solid state of $30{-}50\%$ from an {\it XMM-Newton} RGS spectrum of Sco~X--1. However, \citet{gar11} show that, by taking an improved photoabsorption cross section into
account for atomic oxygen, which itself takes Auger damping into account, the same observation could be adequately modeled without invoking molecular or dust contributions. The best X-ray absorption model should include both atomic and molecular cross sections. However, since the spectral features from solid compounds are expected to be weak, an accurate modeling of the atomic components is a prerequisite.
%-+

\begin{figure}
  \centering
  \includegraphics[width=8.5cm,height=6cm]{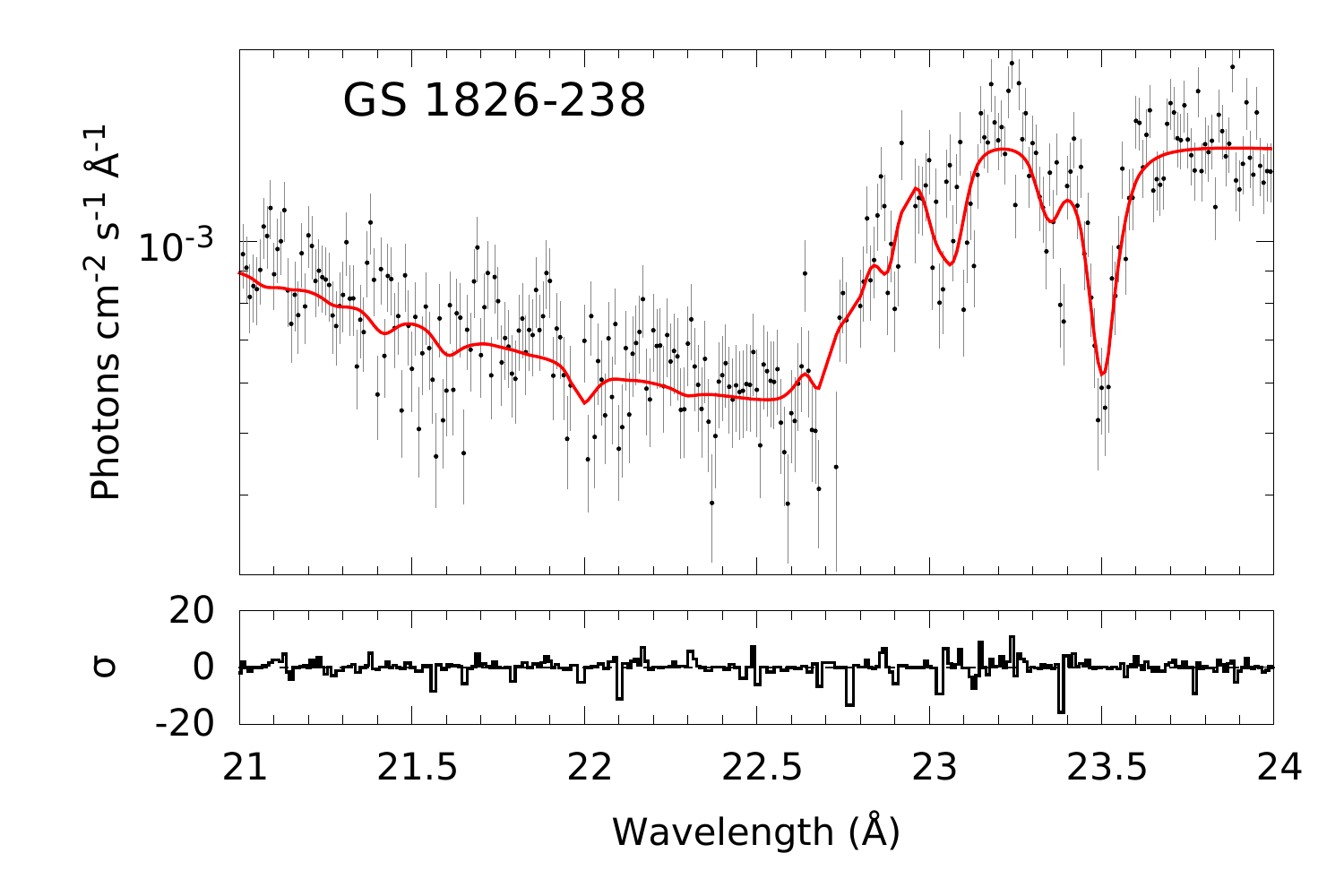}
  \caption{GS~1826--238 {\it XMM-Newton} spectra in the oxygen K-edge region. Observations were combined for illustrative purposes.}\label{fig14}
\end{figure}

In this respect, \citet{pin10} previously analyzed the {\it XMM-Newton} spectra of the binary GS~1826--238, and after modeling the atomic component, they included molecular and dust contributions to diminish the high residuals near the oxygen edge. They derived column densities for andradite (Ca$_{3}$Fe$_{2}$Si$_{3}$O$_{12}$), amorphous ice (H$_{2}$O), carbon monoxide (CO), and hercynite (FeAl$_{2}$O$_{4}$); however, their model included the same undamped atomic cross sections as used in previous studies. Figure~\ref{fig14} shows {\it XMM-Newton} spectra of GS~1826--238 in the 21--24~\AA\ region modeled by {\tt ISMabs}, which considers a more recent and Auger-damped atomic cross section. 

Although {\tt ISMabs} includes neither molecular nor solid photoabsorption cross-sections, it leads to a good fit of the O K-edge absorption features with well-distributed small residuals; that is, the apparent lack of absorption previously perceived by \citet{pin10} was due, in our opinion, to a poor atomic model. Our results imply an upper limit to the amount of oxygen that can be in  a molecular or a solid form.  A quantitative estimate for this limit requires simultaneous fits using both our accurate atomic oxygen K shell cross-sections and the best available models for absorption by molecules and dust. This will be the subject of future investigations.  

It is important to note that, since the valence electronic levels of molecules or dust are more fully occupied than those of atoms, the  K$\alpha$ resonance lines from molecules and dust will in general be weaker than atoms, or absent.  Thus, the addition of molecules or dust to an atomic model when fitting to observed ISM X-ray spectra will lead to a higher ratio of edge to line.  Our fits do not show a systematic discrepancy or statistically significant errors in this ratio, thus reinforcing the conclusion that molecules and grains are a minor contributor to the absorption in the spectral region near the oxygen K edge.
%-+

%%%%%%%%%%%%%%%%%%%%%%%%%%%%%%%%%%%%%%%%%%%%%%%%%%%%%%%%%%%%%%%%%%%%%%%%%%%%%%%

\section{Conclusions}\label{sec_con}

We performed a thorough analysis of the ISM along 24 lines of sight by means of high-resolution X-ray spectroscopy with spectra obtained from the {\it Chandra} and {\it XMM-Newton} space observatories. This is the most complete study to date of this type, and it considered O and Ne charge states. We found that data statistics (i.e., number of counts for a given wavelength region) has a sizeable impact on the analysis, which is particularly important in the case of {\it Chandra} observations taken in CC mode because they require a higher number of counts than those in TE mode in order to avoid background contamination. For sources observed with both telescopes, the derived column densities are in good agreement, so average values are listed.
%-+

Using the {\tt ISMabs} X-ray absorption model, we obtained upper limits for the column densities of \ion{H}{i}, \ion{O}{i}, \ion{O}{ii}, \ion{O}{iii}, \ion{Ne}{i}, \ion{Ne}{ii}, \ion{Ne}{iii}, and metallic Fe. In the case of hydrogen, we found differences between our fits and the 21~cm surveys. These discrepancies may arise from intrinsic absorption variabilities in each source, as well as from the particular choice of continuum representation. Our measurements indicate an increase in the column densities with source distance and a decrease with galactic latitude. We find that the ionization fractions, on the contrary, tend to be constant with distance and latitude reinforcing previous conclusions regarding the dominance of the neutral ISM component. 

We also derived the $N({\rm O})/N({\rm Ne})$ and $N({\rm Fe})/N({\rm Ne})$ abundance ratios finding lower values than solar \citep{gre98}. Using the total column densities, we estimated the O, Ne, and Fe abundances to be $\sim 70\%$, $\sim 87\%$, and $\sim 67\%$,
respectively, relative to the solar standards of \citet{gre98} and $\sim 97\%$, $\sim 122\%$, and Fe~$\sim 67\%$ relative to the revised values of \citet{asp09}.  In general, the abundances show a different trend to the one observed in the column densities: they tend to be constant with distance and latitude. 
%-+

It is important to note that this study only considers a small fraction of the local ISM environment because of the lack of a larger database of bright LMXB to perform a more extensive analysis. Observations with new-generation instrumentation such as {\it Astro-H} will allow a finer examination of the ISM large structures. The present work will be followed by a search and study of molecular and dust absorption features.
%-+

%%%%%%%%%%%%%%%%%%%%%%%%%%%%%%%%%APPENDIX %%%%%%%%%%%%%%%%%%%%%%%%%%%%%%%%%%%%%%%%%%%%%%
\appendix
\label{sec_apx}

\section{Data sample}\label{sec_apx_fit}

For this analysis we used a sample of 84 observations, and the details of the {\it Chandra} and {\it XMM-Newton} observations are listed in Tables~\ref{tab1}--\ref{tab2}. The number of counts for each observation in the broadband fit region (11--24~\AA), O K-edge region (21--24~\AA), Ne K-edge region (13--15~\AA), and Fe L-edge region (16--18~\AA) are also included. In the case of {\it Chandra}, 20 and 29 observations were taken in TE-mode and CC-mode, respectively.
%-+

%%%%%%%%%%%%%%%%%%%%%%%%%%%%%%%%%%%%%%%%%%%%%%%%%
%%%%%%%%%%% observation list %%%%%%%%%%%%%%%%%%%%
%%%%%%%%%%%%%%%%%%%%%%%%%%%%%%%%%%%%%%%%%%%%%%%%%

%%%%%%%%%%% CHANDRA %%%%%%%%%%%
\begin{table*}
\caption{\label{tab1}{\it Chandra} observation list.}
\tiny
\centering
\begin{tabular}{lcccccccc}
\hline
Source   & Galactic &Distance &ObsID  &Mode  &\multicolumn{4}{c}{Counts}\\
 &coordinates &(kpc) &  &&11--24~\AA  &O-edge &Ne-edge &Fe-edge\\
\hline
4U~0614+091     & $(200.87,-3.36)$ & $2.2\pm 0.7^{a}$   &10759&TE&233091&       932&    16561&4084\\
                                &                                  &                               &10760&TE&191848&     752&    13715&3275\\
                                &                                  &                               &10857&TE&346327&     1201&23081&5024\\                               
                                &                                  &                               &10858&TE&121262&     501&    8698&2126\\                             
4U~1636--53     & $(332.9,-4.8)$   & $6.0\pm 0.5^{b}$   &105&TE&244922& 1170&   15732&  3589\\
                                &                                  &                               &1939&TE&212637&      421&    10453&  1984\\
                                &                                  &                               &6635&CC&52613&       408&    3026    &527\\
                                &                                  &                               &6636&CC&209101&      627&    10455&  1507\\
4U~1735--44     & $(346.0,-6.9)$   & $9.4\pm 1.4^{c}$   &704&TE&196485& 1018    &12263& 2828\\
                                &                                  &                               &6637&CC&192055&      651&    10555&  1673\\
                                &                                  &                               &6638&CC&184711&      821&    9370&   1658\\                  
4U~1820--30     & $(2.7, -7.9)$    & $7.6\pm 0.4^{d}$   &1021&TE&113582&        923&    8252    &2160\\
                                &                                  &                               &1022&TE&108264&      1318    &9216&  2966\\
                                &                                  &                               &6633&CC&383259&      2189&   36282&  4991\\
                                &                                  &                               &6634&CC&516960&      2699    &48878& 6502\\
                                &                                  &                               &7032&CC&814603&      4367    &76919& 10493\\         
4U~1957+11      & $(51.30,-9.33 )$ & --                    &4552&TE&168011&     911&    9882    &2415\\
                                &                                  &                               &10659&TE&29828&      116&    1759&   339\\
                                &                                  &                               &10660&TE&36181&      578&    2666    &732\\
Cygnus~X--1     & $(71.3,3.0)$     & $2.10\pm 0.25^{e}$ &1511&CC&247819&        1161    &16914& 3370\\
                                &                                  &                               &2415&CC&890793&      3692    &58878& 13539\\
                                &                                  &                               &3407&CC&1164551&     6437    &65150& 29544\\
                                &                                  &                               &3724&CC&987399&      6665    &45801& 34115\\
                                &                                  &                               &3815&CC&1501955&     4054&   97369&  15254\\
Cygnus~X--2     & $(87.3,-11.3)$   & $13.4\pm 2.0^{c}$  &8170&CC&2156452&       10280&  257811& 30039\\
                                &                                  &                               &8599&CC&2010493&     9237    &229962&        26597\\
EXO~0748--676   &$(279.97,-19.81)$ & $ 8.0\pm 1.2^{c}$  &4573&TE&29582& 883&    2464&   1235\\
                                &                                  &                               &4574&TE&16736&       505&    1300&   681\\
GRO~J1655--40   &$(344.98,2.45)$   &$3.2\pm 0.2^{c}$    &5461&CC&1345632&       2364    &75958& 7893\\
GX~339--4       &$(338.93,-4.32)$  &$10.0\pm 4.5^{f}$   &4420&TE&928061&        647&    50413&  5925\\
                                &                                  &                               &4569&CC&1534889&     1766&   67573&  7565\\
                                &                                  &                               &4570&CC&1561121&     2077&   81294&  10068\\
                                &                                  &                               &4571&CC&1559603&     2084    &86894& 10657\\
GX~349+2        &$(349.10,2.74)$   & $9.2^{g}$          &715&TE&96903   &38     &2643&  184\\
                                &                                  &                               &3354&TE&153152&      104&    3300    &494\\
                                &                                  &                               &6628&CC&152618&      271&    3098&   256\\
                                &                                  &                               &7336&CC&138413&      218     &2922&  244\\                           
                                &                                  &                               &12199&CC&210418&     392&    3904    &364\\
                                &                                  &                               &13221&CC&179170&     876     &5184&  668\\
GX~9+9          &$(8.5,9.0)$       &$4.4^{g}$           &703&TE&226157& 1481&   15013&  3626\\
                                &                                  &                               &11072&TE&831449&     1853&   48170&  7514\\
Sco~X--1        &$(359,23.7)$      &$2.8\pm 0.3^{c}$    &3505&CC&63953& 8914    &232    &16798\\
Ser~X--1        &$(36.11,4.84)$    &$11.1\pm 1.6 ^{c}$  &700&TE&588628& 981&    25989&  4283\\
Swift~J1808-3658                &$(355.38,-8.14)$  &$ 3.61\pm 0.14^{c}$ &6297&CC&82782& 1454&   8500&   2404\\
Swift~J1910.2-0546              &$(29.90,-06.84)$          & --      &  14634&CC&   1218802&        2246& 9806&   74359\\
XTE~J1817--330  &$(359.8,-7.9)$     &$2.5\pm 1.5 ^{h}$  &6615&CC&2622175&       40674&  356751& 96856\\
                                &                                  &                               &6616&CC&2790705&     31440&  300764& 73725\\
                                &                                  &                               &6617&CC&2889136&     20257&  227411& 47640\\
                                &                                  &                               &6618&CC&930503&      9296    &84865& 20114\\
\hline
\end{tabular}
\tablefoot{Distances taken from: $^a$\citet{pae01b}; $^b$\citet{gall08}; $^c$\citet{jon04}; $^d$\citet{kul03}; $^e$\citet{zio05}; $^f$\citet{hyn04}; $^g$\citet{gri02}; and $^h$\citet{sal06}.}
\end{table*}

%--------------------------------------------------------------------------------

%%%%%%%%%%% XMM-NEWTON %%%%%%%%%%%
\begin{table*}
\caption{\label{tab2}{\it XMM-Newton} observation list.}
\tiny
\centering
\begin{tabular}{lcccccccc}
\hline
Source   & Galactic &Distance &ObsID    &\multicolumn{4}{c}{Counts}\\
 &coordinates &(kpc)   &&11--24~\AA  &O-edge &Ne-edge &Fe-edge\\
\hline
4U~0918--54             & $(275.85,-3.84)$ & $5.0\pm 0.7^{a}$    &0061140101&63368&     3194&   15290&  13279\\
4U~1254--69             & $(303.4,-6.4)$   &  $13.0\pm 3.0^{b}$  &0060740101&50859&     1862&   13606&  9325\\  
                                &                             &                                           &0060740901&87761&     3387&   23365&  16126\\ 
                                &                             &                                           &0405510301&205907&    6954&   55922&  37401\\ 
                                &                             &                                           &0405510401&195814&    6879&   53484&  34759\\ 
                                &                             &                                           &0405510501&   193851& 6796&   52792&  34008\\         
4U~1636--53             & $(332.9,-4.8)$   & $6.0\pm 0.5^{c}$    &0500350301&253139&    6259&   71945&  43865\\ 
                                &                             &                                           &0500350401&361332&    8840&   102801& 62238\\ 
                                &                             &                                           &0606070101&219009&    5326&   62231&  37616\\ 
                                &                             &                                           &0606070301&333832&    7815&   95606&  57434\\
4U~1728--16             & $(8.5,9.03)$     & $4.4^{d}$           &0090340101&   233545& 9139&   62514&  42769\\
                                &                             &                                           &0090340601&   469351& 18294&  125676& 86571\\
4U~1735--44             & $(346.0,-6.9)$   & $9.4\pm 1.4^{a}$    &0090340201&334452&    10617&  91311&  59523\\         
                                &                             &                                           &0693490201&   902050& 26482&  245610& 160800\\
4U~1915--05             & $(31.3,-8.4)$    &$8.8\pm 1.3^{a}$            &0085290301&    14733&  404&    4324&   2680\\
Aql~X--1                &$(37.7,-4.1)$     &$5.2\pm 0.8^{a}$            &0303220301&    114415& 2652&   31899&  19293\\
                                &                             &                                           &0303220401&128025&    2971&   35380&  22738\\ 
                                &                             &                                           &0406700201&   688187& 12644&  199459& 113576\\
Cygnus~X--2     & $(87.3,-11.3)$   &$13.4\pm 2.0^{a}$   &0111360101&    587885& 28796&  151920& 113464\\
                                &                             &                                           &0303280101&   1333661& 55636& 342757& 249293\\        
GRO~J1655--40   &$(344.9,2.4)$     &$3.2\pm 0.2^{a}$            &0112921401&    475112& 3326&   138554& 52879\\
                                &                             &                                           &0112921501&   482485& 3266&   139218& 52801\\
                                &                             &                                           &0112921601&516085&    3608&   149411& 57451\\ 
GS~1826--238            &$(9.3,-6.1)$      &$6.7^{c}$                   &0150390101&    345896& 10758&  94316&  65058\\
                                &                             &                                           &0150390301&   304791& 9413&   82648&  57744\\
GX~339--4       &$(338.9,-4.3)$    &$10.0\pm 4.5^{e}$   &0148220201&    337152& 5453&   98860&  54234\\
                                &                             &                                           &0148220301&   259545& 4072&   76761&  40983\\
GX~9+9                          &$(8.5,9.0)$       &$4.4^{d}$                   &0090340101&233545&     9139&   62514&  42769\\
                                &                             &                                           & 0090340601&  469351& 18294&  125676& 86571\\
                                &                             &                                           & 0694860301&  665729& 22912&  180651& 121842\\
Ser~X--1                &$(36.1,4.8)$      &$11.1\pm 1.6^{a}$   &0084020401&    209100& 2904&   60939&  30021\\
                                &                             &                   &0084020501&   199357& 2713&   58235&  28839\\
                                &                             &                                           &0084020601&   206092& 2778&   60447&  29665\\
J1753.5--0127   &$(24.89,12.18)$   &--                          &0311590901&    833709& 3987&   21190&  16690\\
XTE~J1817-330   &$(359.8,-7.9)$    &$2.5\pm 1.5 ^{f}$   &0311590501&    1099719&        20490&  275954& 202834\\
\hline
\end{tabular}
\tablefoot{Distances from $^a$\citet{jon04}; $^b$\citet{int03}; $^c$\citet{gall08}; $^d$\citet{gri02}; $^e$\citet{hyn04}; and $^f$\citet{sal06}.}
\end{table*}

\begin{figure*}
  \centering
  \includegraphics[width=17cm,height=24cm]{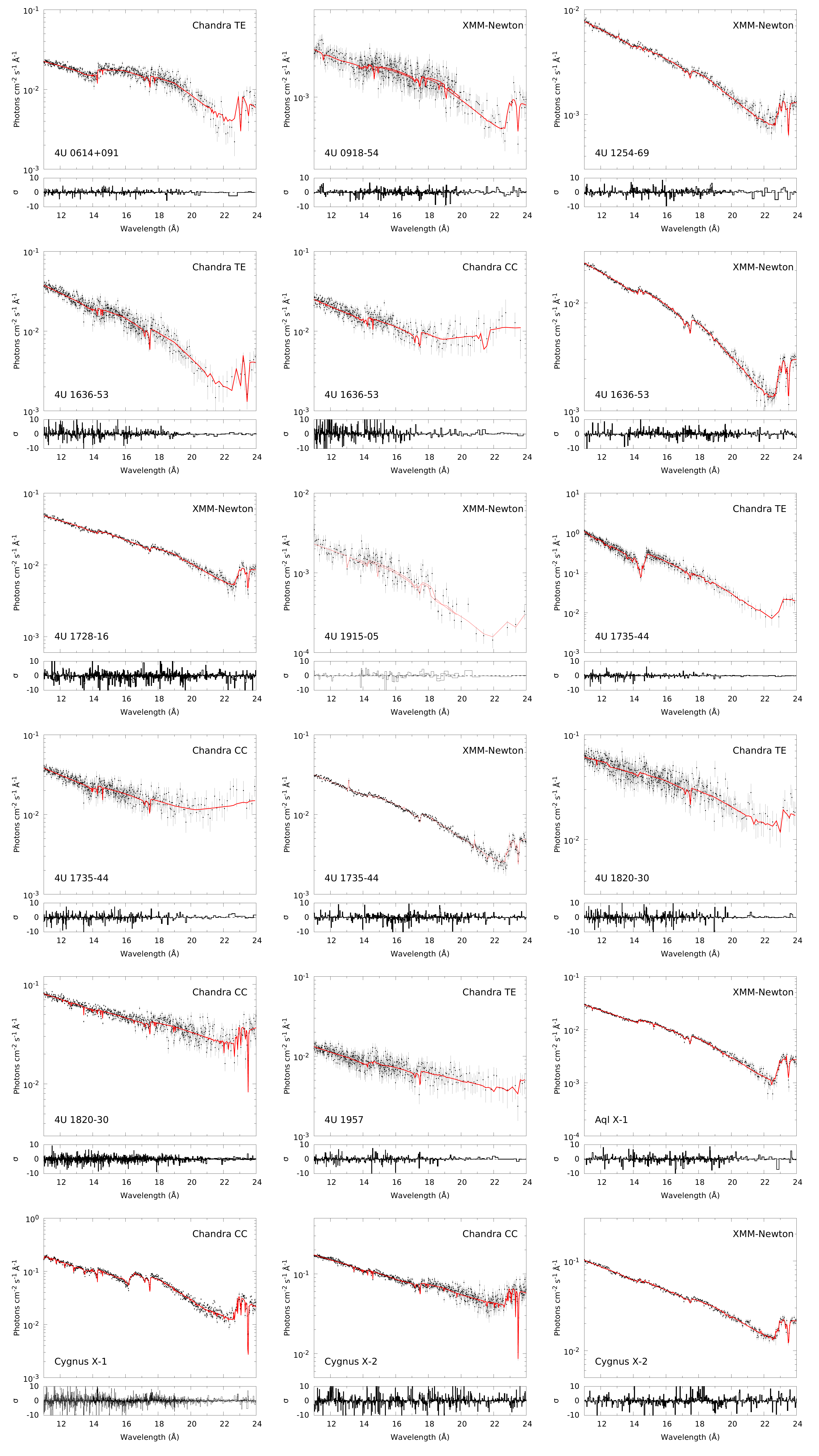}
  \caption{{\tt ISMabs} best broadband fits for the analyzed sources where observations are combined for illustrative purposes. In the case of {\it Chandra} the read mode (CC-mode or TE-mode) is indicated.}\label{fig_fit1}
\end{figure*}

\begin{figure*}
  \centering
  \includegraphics[width=17cm,height=24cm]{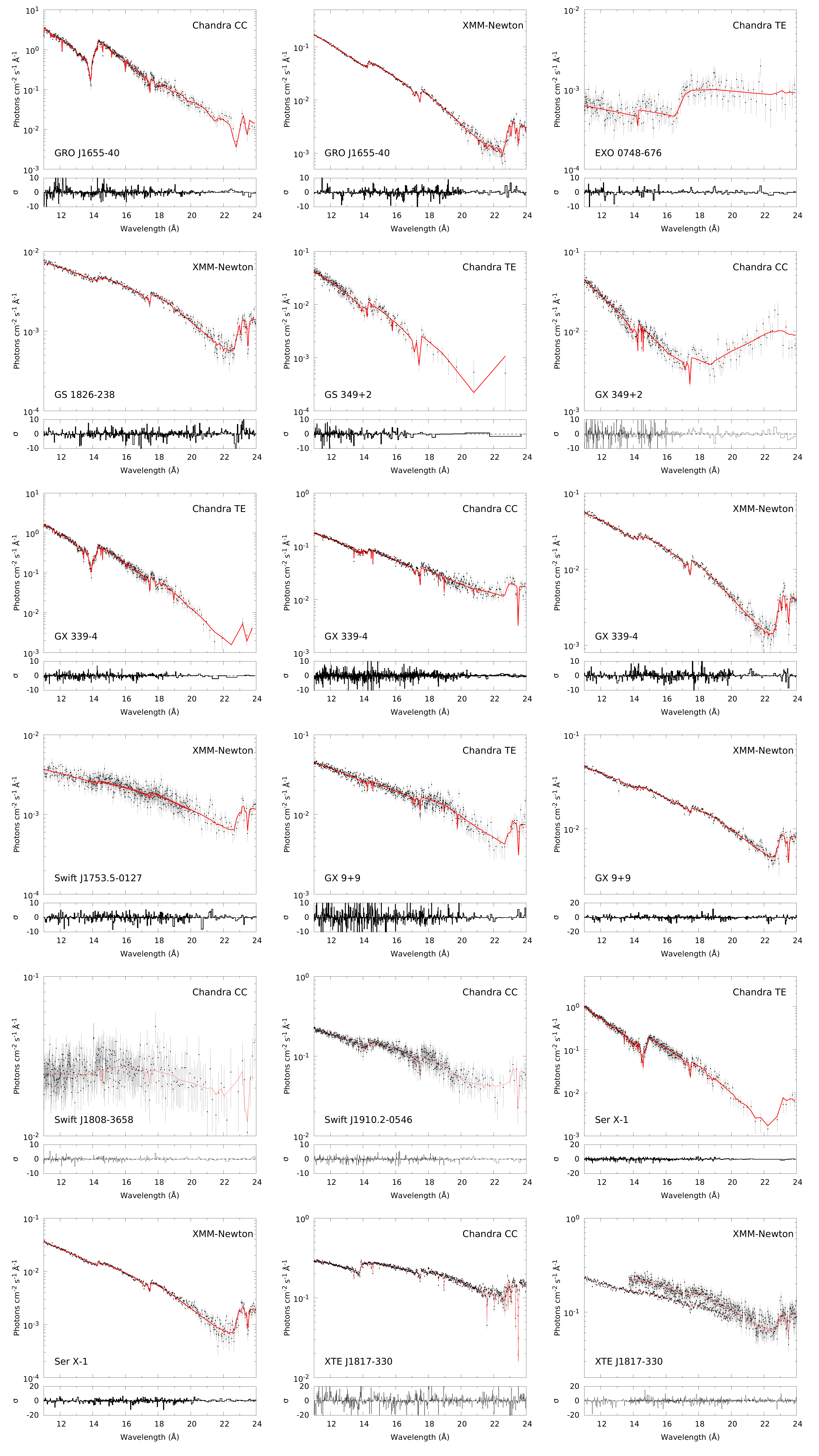}
  \caption{{\tt ISMabs} best broadband fits for the analyzed sources where observations are combined for illustrative purposes. In the case of {\it Chandra,} the read mode (CC-mode or TE-mode) is indicated.}\label{fig_fit2}
\end{figure*}

\begin{figure}
  \centering
  \includegraphics[width=8.5cm,height=6cm]{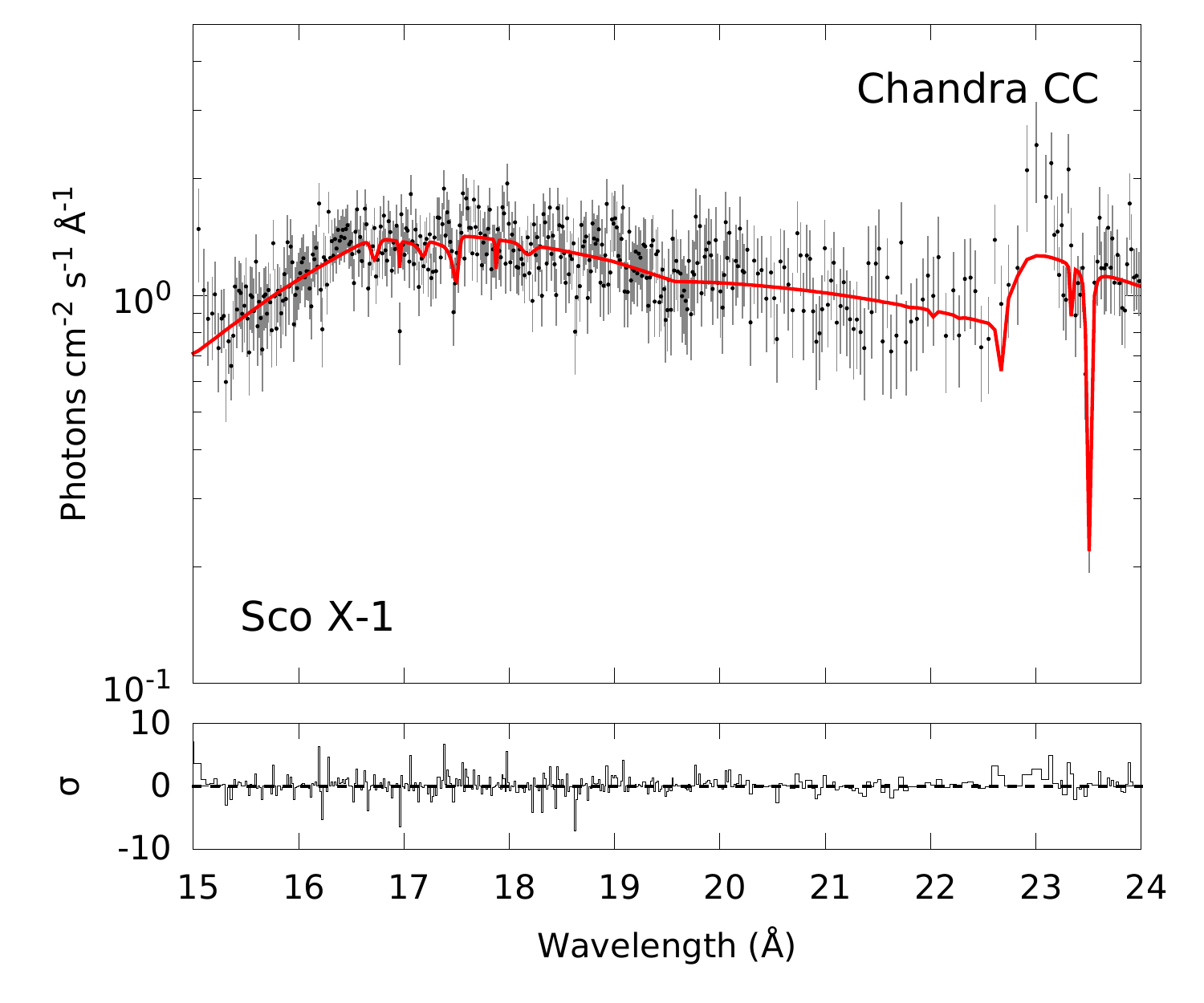}
  \caption{{\tt ISMabs} best broadband fits for Sco~X-1. Observations were combined for illustrative purposes.}\label{fig_fit3}
\end{figure}

Figures~\ref{fig_fit1}--\ref{fig_fit2} show the best broadband fit for each source in flux units. In cases where more than one observation is present, the observations are fit simultaneously, and they vary the continuum parameters so have only been combined for illustrative purposes. The best-fit model is indicated by a solid red line. In the case of Sco~X--1, the broadband fit was performed in the 15--24~\AA\ wavelength range owing to the absence of data below 15~\AA\ (see Figure~\ref{fig_fit3}).  The dominant observed absorption features correspond to the O K edge ($\sim$ 23~\AA), Ne K edge ($\sim$ 14~\AA), Fe L edge ($\sim$ 17~\AA), and the \ion{O}{i} K$\alpha$ transition ($\sim$ 23.5~\AA). In general, a smooth residual distribution (bottom panels) around these edges is observed, indicating modeling accuracy that relies on the latest atomic data. It must be emphasized that the absence of molecular photoabsorption cross-sections in the {\tt ISMabs} model (except
for metallic Fe) does not represent a spectral fitting limitation. This is discussed in Sect.~\ref{sec_apx_mol}.
%-+
All data were rebinned to 25 counts per channel and combined for illustrative purposes

%--------------------------------------------------------------------------

%--------------------------------------------------------------------------

%--------------------------------------------------------------------------

\section{{\it Chandra} CC-mode notes}\label{sec_apx_cc}

The {\it Chandra} CC-mode provides a fast readout mode to decrease the pileup effect in bright sources: i.e., the detection of two photons as a single event with the sum of their energies. While in TE-mode the events are collected for a specific time frame and then read out collectively, in CC-mode they are read out continuously by collapsing them to one row. In the case of a low-count regime, CC-mode data must be analyzed carefully because they can be affected by the background that cannot be separated from the source spectrum.
%-+

\begin{figure*}
  \resizebox{\hsize}{!}{\includegraphics{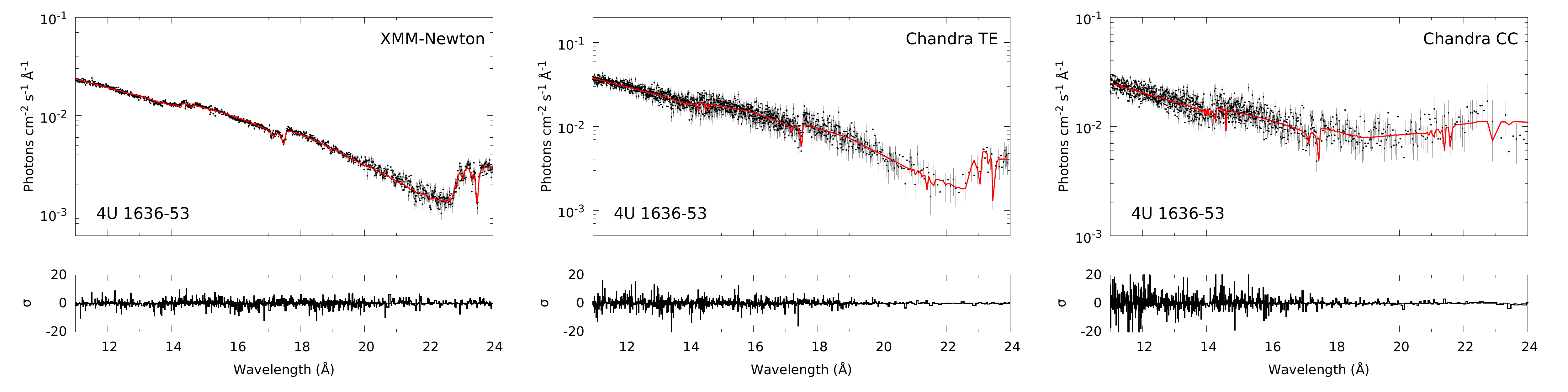}}
  \caption{Comparison of the 4U~1636--53 spectra obtained with {\it XMM-Newton}, {\it Chandra} TE-mode, and {\it Chandra} CC-mode in the broadband region. Observations are combined for illustrative purposes.}\label{fig13}
\end{figure*}

As an example of this effect, we show a comparison in Figure~\ref{fig13} of the 4U~1636--53 spectra in the 11--24~\AA\ region as obtained by the {\it Chandra} TE-mode, CC-mode, and {\it XMM-Newton}. All data have been rebinned to 25 counts per channel and combined for illustrative purposes. The total number of counts in the oxygen absorption region (21--24~\AA) without rebinning is 1591 (CC-mode), 1035 (TE-mode), and 28240 (RGS). The \ion{O}{i} and \ion{O}{ii} K$\alpha$ absorption lines at $\sim 23.5$~\AA\ and $\sim 23.35$~\AA, respectively, are clearly observed in the TE-mode and the RGS data, but they cannot be detected in CC-mode due to background contamination that becomes dominant at $\sim 18{-}20$~\AA. For this reason and based on previous tests, we discarded from our analyses the oxygen column densities derived from observations in CC mode with fewer than 2000 counts in the oxygen K-shell region.
%-+

%--------------------------------------------------------------------------

\bibliographystyle{aa}

% WARNING
%-------------------------------------------------------------------
% Please note that we have included the references to the file aa.dem in
% order to compile it, but we ask you to:
%
% - use BibTeX with the regular commands:
%   \bibliographystyle{aa} % style aa.bst
%   \bibliography{Yourfile} % your references Yourfile.bib
%
% - join the .bib files when you upload your source files
%-------------------------------------------------------------------

\end{document}